# Impact of attractive interactions on the rheology of dense athermal particles

Ehsan Irani,[1] Pinaki Chaudhuri,[2] and Claus Heussinger[1]

[1]*Institute for Theoretical Physics, Georg-August University of Göttingen,*
*Friedrich-Hund Platz 1, 37077 Göttingen, Germany*
[2]*Institut für Theoretische Physik II, Heinrich-Heine-Universität Düsseldorf, 40225 Düsseldorf, Germany*
(Dated: May 20, 2014)

Using numerical simulations, the rheological response of an athermal assembly of soft particles with tunable attractive interactions is studied in the vicinity of jamming. At small attractions, a fragile solid develops and a finite yield stress is measured. Moreover, the measured flow curves have unstable regimes, which lead to persistent shearbanding. These features are rationalized by establishing a link between the rheology and the inter-particle connectivity, which also provides a minimal model to describe the flow curves.

In nature, soft disordered solids occur in different forms (eg. gels, emulsions, colloids, foams, grains etc) across a wide range of packing fractions $\phi$, which is made possible by the tuning of particle interactions. The flow properties of these soft materials have been harnessed for various applications, e.g. in the food or the chemical industry. Thus, understanding the role of particle interactions and the corresponding mechanisms which lead to observed rheological behaviour is an important recurrent theme.

For non-Brownian suspensions of frictionless repulsive spheres, it is observed that ramping up the packing fraction results in the occurrence of jamming at $\phi = \phi_J$ [1, 2]. The rheological signature of the onset of jamming is the development of a finite yield stress at $\phi_J$ [3]. For *Brownian* suspensions of such particles, it has been shown that a yield stress exists at $\phi < \phi_J$, due to the presence of thermal vibrations [4]. A similar systematic investigation of how the jamming paradigm is changing upon the introduction of attractive particle interactions is still missing. It is known that at smaller $\phi$, such systems do exhibit finite yield stress [5, 9, 19, 24]; but, a quantitative bridge with the jamming scenario needs to be developed.

Such studies are also needed since shear-banding, the phenomenon of spatially inhomogeneous flows observed in many soft yield-stress fluids [6, 7], has often been attributed to attractive interactions [8, 9]. In general, persistent occurrence of shear-bands has been linked to non-montonic constitutive laws leading to flow instabilities [6, 10] (e.g. in micelles [11]). It is not known how interparticle attractions could result in such instabilities.

Conceptually, one can imagine the steady flowing state to be a regime where there is a continuous competition between the rupturing induced by shear and processes that try to restore local structure. Theoretical models suggest that non-monotonic flow curves can occur due to long-lived local fluidizations when the post-rupture restructuring takes a very long time [12–14]. However, experiments and numerical simulations have shown that for $\phi > \phi_J$, no such instabilities occur in the flow curve for either repulsive or attractive systems [15–18]. The question now arises whether for $\phi < \phi_J$, a short-ranged attraction which introduces a new lengthscale for structure formation leads to longer restoration timescales and if this is indeed the origin of a shear banding instability.

In this letter, we report a simulational study of the variation in rheological behaviour of an athermal assembly of soft disks, near $\phi_J$, by the tuning up of attractive interactions. We show that for $\phi < \phi_J$, minimal attractions result in finite yield stresses. The variation of this threshold with attraction and packing fraction can be rationalised in terms of changing structure, viz. the number of contacts per particle and its link with isostaticity. Further, we demonstrate for the first time the existence of non-monotonic flow curves at these weak attractions, causing persistent shearbanding over a range of shear-rates. Thus, our work reveals new rheological behaviour in the vicinity of $\phi_J$ with the introduction of attractive interactions and demonstrates how the flow properties gradually deviate from that of repulsive particles.

In our numerical simulations (using LAMMPS [20]), we study a two-dimensional 50:50 binary mixture of soft disks, having a size ratio of 1.4. The disks interact via the following potential, which can be considered to be a model for cohesive grains or attractive emulsions [18, 19]:

$$V(r_{ij}) = \begin{cases} \epsilon\left[(1-\frac{r_{ij}}{d_{ij}})^2 - 2u^2\right], & \frac{r_{ij}}{d_{ij}} < 1+u \\ -\epsilon\left[1+2u-\frac{r_{ij}}{d_{ij}}\right]^2, & 1+u < \frac{r_{ij}}{d_{ij}} < 1+2u \\ 0, & \frac{r_{ij}}{d_{ij}} > 1+2u \end{cases} \quad (1)$$

where $d_{ij} = (d_i + d_j)/2$, and $d_i$ being the diameter of disk $i$. Such a potential results in piecewise-linear interaction forces. The strength and the range of the attractive forces are simultaneously tuned by varying $u$ (see inset of Fig. 2 for a schematic). We shear the system of particles at any imposed shear-rate $\dot{\gamma}$ by using the appropriate Lees-Edwards boundary conditions. During the flow, when two particles overlap, they experience a dissipative force which depends on their relative velocity: $-b[(\vec{v}_i - \vec{v}_j).\hat{r}_{ij}]\hat{r}_{ij}$, where $b$ is the damping coefficient, and

$\hat{r}_{ij}$ is the unit vector between particles $i$ and $j$. We integrate the corresponding Newton's equations of motions for different system sizes $N = 10^3, 10^4, 2 \times 10^4$ in order to explore the rheological properties for a wide range of packing fractions $\phi$. In our simulations, the units for energy, length and time are respectively $\epsilon$, $d_s$ and $\sqrt{md_s^2/\epsilon}$, where $m$ is the mass of the particles and $d_s$ is the diameter of the smaller particles. Further, by our choices of $m=1$, $\epsilon=1$, $b=2$, the system of particles undergo overdamped dynamics via inelastic collisions [22, 23].

First, we focus on how the flow curves ($\sigma$ vs $\dot{\gamma}$) shape up after the attractive interactions are introduced. In Fig. 1(a), for a system size of $N = 10^3$, we show the flow curves at $\phi = 0.82$ (which is less than $\phi_J \approx 0.843$). For the purely repulsive system ($u = 0$) we observe the usual Bagnold scaling $\sigma \sim \dot{\gamma}^2$ [23]. As soon as the attraction strength is finite, the particle assembly exhibits a finite (albeit small) yield stress $\sigma_y$. The yield stress increases with increasing attraction, which is expected. We also note that at larger $\dot{\gamma}$, the flow curves for all attraction strengths collapse and are identical to the repulsive case. Thus attraction has an effect only at small shear-rates and the range over which this change occurs increases with increasing attraction strength. We will call these two regimes "attraction-dominated" and "repulsion-dominated" flow in the following.

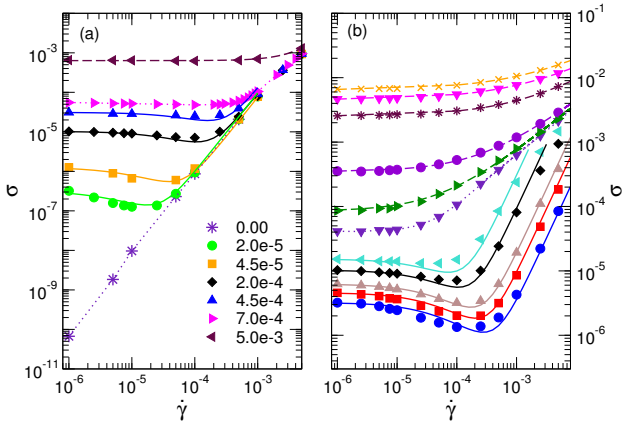

FIG. 1: Shear stress $\sigma$ as function of strainrate $\dot{\gamma}$ for $N = 10^3$: (left) for different attraction strengths $u$ (as specified in the legends) at $\phi = 0.82$; (right) for different packing fractions (from the bottom to top) $\phi = 0.75, 0.78, 0.80, 0.82, 0.83, 0.84, 0.843, 0.85, 0.9, 0.95, 1.0$ for the attraction strength $u = 2 \cdot 10^{-4}$. The solid lines are fits using the fluidity model (Eq. 3), the dashed lines are Herschel-Bulkley fits and the dotted lines are guides to the eye.

In Fig. 1(b), we show the variation of the flow curves with packing fraction $\phi$ for a fixed $u = 2 \cdot 10^{-4}$. We observe that the system exhibits a finite $\sigma_y$ at $\phi$ much below $\phi_J (\approx 0.843)$. At $\phi > \phi_J$ we observe the usual Herschel-Bulkley form, consistent with previous work [18].

In both panels of Fig. 1, the flow curves are visibly non-monotonic, for either (a) low attraction strengths or (b) low packing fractions. In both cases, there exists an intermediate regime of shear rates, where shear stress is a decreasing function of strain rate $\dot{\gamma}$. As discussed earlier, such flow curves lead to localized shear bands, i.e homogeneous flow is no longer possible.

It is known that under imposed $\dot{\gamma}$, shear band formation can be avoided if the wavelengths of the unstable modes do not fit into the lateral size of the simulation box [21]. Thus, in our simulations, for a system size of $N = 10^3$ (data in Figs. 1-4), velocity profiles measured in the unstable regime of the flowcurve are seen to be linear, i.e homogeneous flow is observed. However, when the system size is increased to $N = 10^4, 2 \times 10^4$, the rapid formation of permanent shear bands, in this regime, is observed. Further, in this unstable part, when stress-controlled simulations are done, we observe either runaway flow towards the stable high-shear rate branch or absorption into an arrested state [27].

With the minimum in the flow-curve being quite shallow, the tendency to form shear bands in our system is weak. This gives us a rare opportunity to study not only the properties and formation of shear bands but also the underlying, nominally unstable, constitutive law.

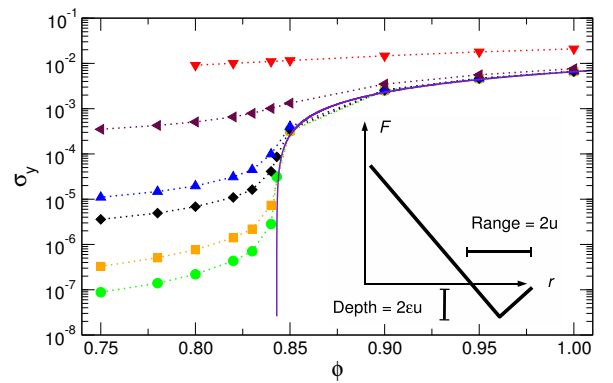

FIG. 2: Variation of yield stress $\sigma_y$ with packing fraction $\phi$ for different attraction strengths (from bottom to top) $u = 2 \cdot 10^{-5}, 4.5 \cdot 10^{-5}, 2 \cdot 10^{-4}, 4.5 \cdot 10^{-4}, 5 \cdot 10^{-3}, 5 \cdot 10^{-2}$. Solid line: yield stress of the repulsive system ($u \to 0$) is expected to vanish at $\phi = \phi_J$ as $\sigma_y^{(\text{rep})} \propto (\phi - \phi_J)^\alpha$; $\alpha = 1.04$. (Inset): schematic of the particle interaction force $F(r)$.

By gathering data for different $\phi$ and $u$, we look at the variation of the yield stress, $\sigma_y \equiv \sigma(\dot{\gamma} \to 0)$, which is estimated from the stress at the smallest available strainrate ($\dot{\gamma} = 10^{-6}$ or $10^{-7}$); this is shown in Fig. 2. For high volume-fractions $\phi$ and small attraction strength $u$ – in the repulsion-dominated regime – the yield stress is independent of $u$ and scales as $\sigma_y^{(\text{rep})} \propto (\phi - \phi_J)^\alpha$. The fitted value of the exponent $\alpha \approx 1.04$ is consistent with previous results for purely repulsive particles [3], but likely suffers from finite-size effects [25]. For strong attraction the yield stress is only weakly density dependent and scales linearly with the strength of attraction, i.e. $\sigma_y \sim u$. Such a property is trivially expected from the rupture of a single element of strength $u$.

The new and non-trivial result is the regime at small $u$ and below the jamming limit ($\phi < \phi_J$). There, a finite yield-stress is observed even at densities nominally far below $\phi_J$, where the corresponding repulsive system is a normal fluid. Thus, in weakly attractive systems, the cross-over from attraction-dominated to repulsion-dominated flow can also be observed by increasing $\phi$. Note the similarity with the repulsive but Brownian system [4], where a crossover occurs from a "weakly thermal" regime ($\phi < \phi_J$) to an athermal regime ($\phi > \phi_J$).

What is the proper energy scale in the weakly attractive regime? In order to answer this question, we first take a look at the connectivity $z$, defined as the average number of contacts per particle. In counting the contact number, we include all nearest neighbours lying within the range of the interaction potential. The typical variation of $z$ as a function of strain rate is shown in the two panels of Fig. 3 for (a) fixed $\phi$ and (b) fixed $u$; the values of the different parameters are the same as in Fig. 1.

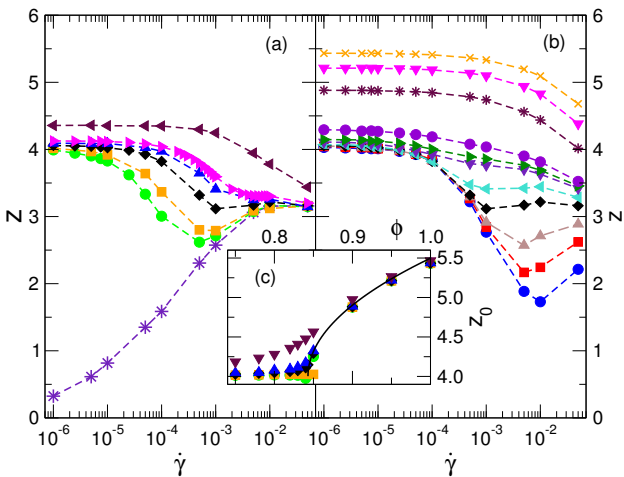

FIG. 3: Variation of coordination number $z$ as a function of strainrate $\dot\gamma$ - (a) for different attraction strengths ($u$) at $\phi = 0.82$ and (b) for different $\phi$ at a fixed $u = 2 \cdot 10^{-4}$. The values of $u$ in (a) and $\phi$ in (b) are the same as those in Fig.1. Dashed lines are guides to the eye. (c) Variation of $z_0 \equiv z(\dot\gamma \to 0)$ with $\phi$. The solid line denotes $(\phi - \phi_J)^{1/2}$.

We concentrate on small strain rates first. For the repulsive particles, $z \to 0$ at vanishing shear-rates, as expected for our model of inertial, dissipative dynamics [23]. However, as soon as $u$ is finite, $z_0 \equiv z(\dot\gamma \to 0)$ jumps to a finite value. In both the panels of Fig. 3, we notice that $z_0$ saturates at a value not much larger than the isostatic limit $z_{\rm iso} = 4$ as $\dot\gamma \to 0$. Thus, minimal attraction leads to similar isostatic structures as seen in the purely repulsive system for $\phi = \phi_J$. The difference being that, here, isostatic networks are observed over a range of volume fractions and considerably below $\phi_J$. At high $\phi$ the familiar scaling law, $z_0 - z_{\rm iso} = \zeta_0(\phi - \phi_J)^{1/2}$; $\zeta_0 \approx 3.78$ [26] is recovered (see Fig.3(c)).

Previous work on packings of soft repulsive particles and elastic networks [1] have shown how linear elasticity in the near-isostatic regime can be understood in terms of the deviation from isostaticity, $\delta z = z_0 - z_{\rm iso}$. It turns out that we can use a similar reasoning to derive the scaling form for the yield stress in the attraction-dominated regime to be $\sigma_y^{\rm (att)} \sim u^{1/2}\delta z^{3/2}$ (see Supp. Mat. [27]). With this and the repulsive yield stress $\sigma_y^{\rm (rep)} \sim \delta\phi^\alpha \theta(\delta\phi)$ the overall yield stress can be written as follows

$$\sigma_y/|\delta\phi|^\alpha \propto \begin{cases} u^{1/2}\delta z^{3/2}/|\delta\phi|^\alpha, & \sigma_y^{\rm (rep)} \ll \sigma_y^{\rm (att)} \\ 1, & \sigma_y^{\rm (rep)} \gg \sigma_y^{\rm (att)} \end{cases}. \quad (2)$$

This scaling form is verified in Fig. 4(a), using the data for $\sigma_y$ shown in Fig. 2 and the corresponding data for $z_0$. The data collapse on the two branches defined by Eq. (2) is excellent and holds over several orders of magnitude. We also note some deviations for the smallest attraction strengths and packing fractions.

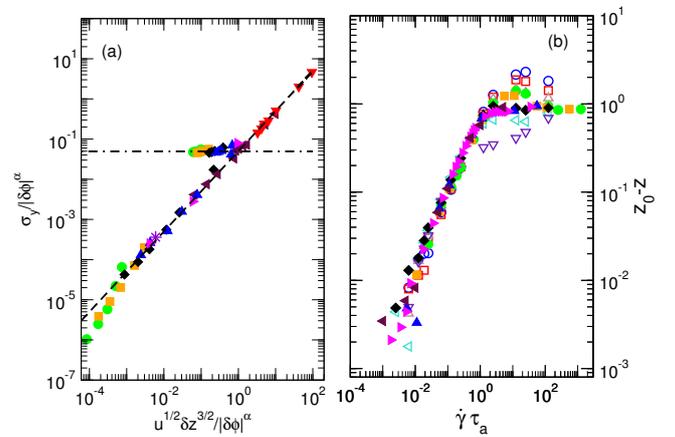

FIG. 4: *Scaling plots*: (a) Yield stress $\sigma_y$ normalized by $|\delta\phi|^\alpha$ as a function of the combination $u^{1/2}\delta z^{3/2}|\delta\phi|^{-\alpha}$. Dashed lines correspond to the two regimes of Eq. (2). (b) Variation of $z_0 - z$ with $\dot\gamma\tau_a$ (where $\tau_a = 0.5/u$), using data shown in Fig. 3(a) (closed symbols) and Fig. 3(b) (open symbols).

We return to discussing the rheology at finite $\dot\gamma$, where a similar link exists. As with the non-monotonic flow curves, we see a non-monotonic behaviour also in $z$ vs. $\dot\gamma$ (Fig. 3). At small strainrates, where $\sigma(\dot\gamma)$ is decreasing, $z(\dot\gamma)$ quickly drops to values far below $z_{\rm iso} = 4$, before it rises again following the repulsive branch.

Here, we see the manifestation of the two competing mechanisms described in the introduction: shear-induced rupture of the fragile near-isostatic network and attraction-induced aggregation (see the supporting movies in the Supp. Mat. [27]). At small but finite $\dot\gamma$, the imposed shear is not fast enough to efficiently destroy the ever continuous restructuration. At large $\dot\gamma$, on the other hand, the intrinsic relaxation time is too large to lead to the built-up of a large aggregate.

To extract a characteristic time-scale for this aggregation process, we demonstrate that by using the scaling form $z(\dot{\gamma}) = z_0 - \dot{\gamma}\tau_a$, it is possible to collapse all the data for $z$ vs $\dot{\gamma}$ in the regime of weak attraction (see Fig. 4(b)). While generating the scaling collapse, we obtain $\tau_a \approx 0.5/u$, for the intrinsic timescale for restructuration. Thus, for weak attractions, $\tau_a$ is large. Now, for shearbanding to occur, applied shear-rates need to satisfy $\dot{\gamma}\tau_a < 1$. Hence, shearbanding can only be observed in the regime of small $u$, which agrees with the flow-curves of Fig. 1.

Moreover, we can use the attraction-dependent time-scale within a simple model to provide a reasonable fit to these flow-curves. Based on the *fluidity* approach of Picard *et al.* [28], we can derive (for details, see Supp. Mat. [27]) the following simple expression for the stress

$$\sigma(\dot{\gamma}) = \sigma_y^{(\text{att})} \cdot \frac{W(\dot{\gamma}\tau)}{\dot{\gamma}\tau} + \sigma^{(\text{rep})}(\dot{\gamma}). \quad (3)$$

$W(x)$ is the Lambert-$W$ function, and the time-scale is taken to inversly depend on attraction strength $\tau \sim u^{-1}$ (i.e. proportional to $\tau_a$). The repulsive branch is assumed to show Bagnold-scaling, $\sigma^{(\text{rep})} \sim \dot{\gamma}^2$. The underlying physics of the model is the above mentioned competition between shear-induced fluidization and intrinsic relaxation/aggregation. Despite the simplicity of the model, the non-mononotic flow-curves can be fitted surprisingly well, as shown in Fig. 1(a)-(b). Nevertheless, the model cannot reproduce some details of the simulation data: for example, the precise functional form in the limit of small $\dot{\gamma}$. The fluidity model, (as well as others [13, 14, 29]) give $\sigma \to \sigma_y/(1+\dot{\gamma}\tau)$ for small $\dot{\gamma}$. The simulation data hints at a weaker (logarithmic) dependence on strainrate in this limit. More work is needed to resolve this issue, both from a theoretical point of view as well as from the simulations.

The link between connectivity and flow is further illustrated when, for a shearbanded state, one measures the spatial profiles of *local* shear rates and the corresponding *local* connectivity. This is shown in Fig. 5, at a state point $\{\phi, \dot{\gamma}, u\}$ in the unstable regime of the flow curve (see Fig. 1(a)) for a large enough system size and measured during a long strain window. It is clear that the flowing region has a low connectivity, while the arrested band is nearly isostatic with $z \approx 4$. Future studies should address formation and properties of these shear bands.

To conclude, in the proximity of $\phi_J$, we have studied how weak attractive interactions ($u$) change the rheological properties of dense disordered assemblies of non-Brownian particles. First, we rationalized the existence of finite yield stresses below the (repulsive) jamming transition via a scaling argument that exploits the near-isostatic nature of the contact network, viz. $\sigma_y \sim u^{1/2}(z-z_{\text{iso}})^{3/2}$. Secondly, we demonstrated the occurrence of non-monotonic flow curves indicating a shear banding instability. We showed that this feature is a consequence of a long structural aggregation time-scale $\tau \sim u^{-1}$, which can be extracted from the loss of connectivity as the shear rate is increased. With this time-scale at hand, we set up a fluidity model to provide reasonable fits to the non-monotonic flow-curves. Thus, we established how the emerging rheological changes are linked to properties of the contact network.

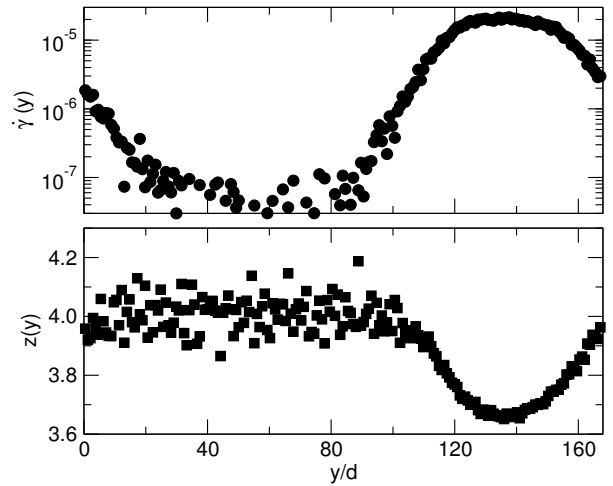

FIG. 5: At $\phi = 0.82$, $u = 2 \cdot 10^{-5}$ and $\dot{\gamma} = 5 \times 10^{-6}$: Spatial profile of (top) local shear rate $\dot{\gamma}(y)$ and (bottom) the corresponding local contact numbers $z(y)$ for $N = 2 \times 10^4$.

An expected consequence of the non-monotonicity is that static and dynamic yield stresses will be different (as, e.g., reported in Ref. [24]). However, the inverse does not necessarily follow, i.e. a difference between static and dynamic thresholds does not necessarily imply a non-monotonic flow-curve. Thus, independent studies using imposed stress and imposed strain rate are necessary.

Future work should explore the impact of thermal fluctuations on the rheological behaviour observed by us, thus making the possible link with the flow behaviour of dense gel-glasses. Also, studies should be extended to lower packing fractions where more open-ended fragile networks of the attractive particles are expected to occur [19, 30]. In parallel, systematic experiments are necessary at these packing fractions in order to further test our findings. While there have been recent experiments probing static properties of jammed attractive assemblies [31] or their shear moduli [32], more detailed rheological studies of these dense fragile networks are necessary.

We acknowledge financial support by the DFG via the Emmy Noether program (He 6322/1-1). We also thank J. Horbach for useful discussions.

---

# Supplemental material
# Impact of attractive interactions on the rheology of dense athermal particles

Ehsan Irani,[1] Pinaki Chaudhuri,[2] and Claus Heussinger[1]

[1]*Institute for Theoretical Physics, Georg-August University of Göttingen,
Friedrich-Hund Platz 1, 37077 Göttingen, Germany*
[2]*Institut für Theoretische Physik II, Heinrich-Heine-Universität Düsseldorf, 40225 Düsseldorf, Germany*

(Dated: April 22, 2014)

## Movies

The movies illustrate the qualitative difference between the attraction-dominated flow at small strainrates and repulsion-dominated flow at large strainrates. Volume fraction is $\phi = 0.75$, attraction strength $u = 2 \cdot 10^{-5}$.

1. attraction.dominated.avi
   Strainrate $\dot{\gamma} = 10^{-6}$; attraction-induced aggregation of an isostatic network that breaks and reforms under shear

2. repulsion.dominated.avi
   Strainrate $\dot{\gamma} = 10^{-4}$; repulsion-dominated flow, where attractive forces play no role. particle motion as a sequence of flow-induced collisions.

## Flow under imposed stress

Similar to earlier studies [1], we consider flow under imposed stress by confining the system of particles between rough walls and then imposing the external stress by pulling the top wall with a constant force. The applied force ($F_0$) is determined by the desired stress ($\sigma_0$) : $F_0 = \sigma_0 L_x$, where $L_x$ is the length of the confining boundary. The interaction potential between the wall particles and the neighbouring fluid particles is chosen to be the same as that of those in the bulk.

In Fig.1, we show data for a system of $N = 2916$ particles at $\phi = 0.825$ and $u = 5 \times 10^{-5}$. The non-monotonic flow curve for these parameters is shown in the inset of Fig.1. Using a configuration from a steady state flow under imposed strain rate of $10^{-6}$ (which is in the unstable branch of the flow curve), we switch to a stress-controlled run with an imposed stress of $\sigma_0 = 1.593 \times 10^{-6}$; the corresponding strain rate ($\dot{\gamma}_w$) as monitored at the top wall is shown in Fig.1. We observe that rather than flowing at the initial shear-rate of $10^{-6}$, the flow rapidly evolves to an eventual flowing state of $\dot{\gamma}_w = 3.5 \times 10^{-5}$ (which is the corresponding state on the stable branch of the flow curve; see inset of Fig.1). We also report that for smaller imposed stresses, we observe that the system eventually evolves to an arrested state and the flow stops.

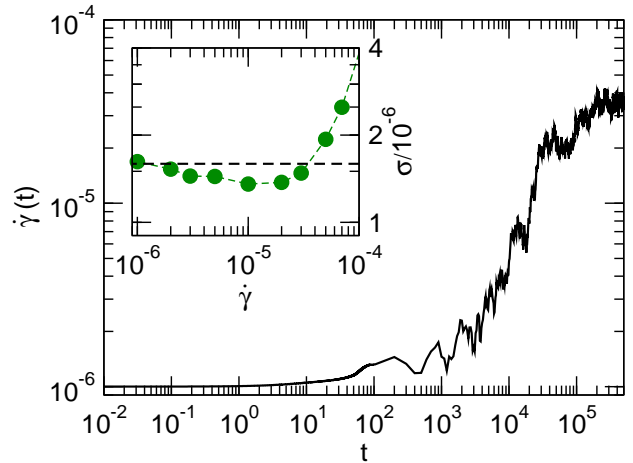

FIG. 1: (Inset) Flow curve for $\phi = 0.825, u = 5 \times 10^{-5}$, measured from strain-rate controlled simulations. The dashed line corresponds to the imposed stress $\sigma_0/10^{-6} = 1.593$ in the stress controlled simulation. (Main) Under the applied stress, time evolution of strain rate as measured at the wall : $\dot{\gamma}_w = v_w/\ell_y$; where $v_w$ is the measured wall velocity and $\ell_y = 49.41$ is the distance between the confining rough walls.

## Yield stress

Here, we present details for the derivation of the relation $\sigma_y \sim u^{1/2} \delta z^{3/2}$.

Starting point of the derivation is the linear response to a small shear strain $\gamma$.

In elastic spring networks and systems of soft repulsive particles, the linear response near isostaticity ($\delta z \to 0$) is characterized by strong non-affine motion [2], quantified by relative particle displacements $\delta_\perp \sim \gamma/\delta z^{1/2}$, that are directed tangentially to the particle contact. Close to the isostatic point, such a response is energetically less costly as an affine deformation, which would lead to particles pressing into each other. The associated linear-elastic shear modulus is known to be $g_{\text{lin}} \sim \delta z$ [2].

These scaling laws are known to hold for systems of soft repulsive particles as well as networks of elastic springs. In our system, as long as the particle contact is not broken, the inter-particle force just behaves like a harmonic spring with a range set by the attraction $u$. Thus, for motion amplitudes smaller than this range our system should behave just like a network of elastic springs.

Yielding can then be defined as the point where the motion amplitude exceeds the range of the attraction.

To determine this point, we consider non-linear loading conditions, where the tangential particle motion also leads to higher-order longitudinal contributions, $\delta_\parallel \sim \delta_\perp^2/a$ (Pythagoras) [3, 4]. In the weakly attractive particle system that we study, the maximal dilational strain is set by the range of the attractive potential $\delta_\parallel < u$. This gives for the yield strain $\gamma_y \sim (u\delta z)^{1/2}$, and for the yield stress $\sigma_y \sim g_{\text{lin}}\gamma_y \sim u^{1/2}\delta z^{3/2}$.

### Fluidity model

Here, we describe details of the fluidity model presented in the manuscript. Common starting point of many different models in this context [5] is to split the stress into a "viscous" term and an "elastic" contribution from the microstructure:

$$\sigma = \eta\dot{\gamma} + \sigma_s \quad (1)$$

The rheological properties then follow from the temporal evolution of $\sigma_s$, which is provided by an additional equation of motion.

Specific to our system, we take the viscosity $\eta \sim \dot{\gamma}$ to comply with the Bagnold scaling in the fluid branch. In the context of the fluidity model, elastic stresses are described via the evolution of the fluidity $a$: $\sigma_s = \dot{\gamma}/a$ [6]. Being an inverse viscosity, the fluidity is a measure for the extent of aggregation into a network structure. High fluidity implies little aggregation, while vanishing fluidity corresponds to a solid-like state (finite $\sigma_s$ but $\dot{\gamma} = 0$). Different choices for the evolution equation for $a$ are possible, as mentioned in Ref. [6]. Here, we take a simple exponential relaxation

$$\dot{a} = -a/\tau_0 + r_1\sigma_s\dot{\gamma} \quad (2)$$

with a relaxation rate $1/\tau_0 = e^{-r_0\sigma_s\dot{\gamma}}$. These ingredients reflect the above mentioned physics of shear-induced fluidization ($r_1\sigma_s\dot{\gamma}$) that is counterbalanced by an intrinsic relaxation process. Crucially, this relaxation rate itself depends on the shear [6], such that shear also inhibits relaxation into a less-fluid state. In our system this is warranted, because local energy input $\sim \sigma_s\dot{\gamma}$ needs to be dissipated (via inelastic collisions) before a rigid state is formed. Solving these equations in steady-state one arrives at

$$\sigma_s = \sigma_y \frac{W(\dot{\gamma}\tau)}{\dot{\gamma}\tau} \quad (3)$$

with the Lambert-$W$ function $W(x)$, the yield stress $\sigma_y = 1/\sqrt{r_1}$ and the time-scale $\tau = r_0/2\sqrt{r_1}$. Thus, we obtain the model that is used in the main text for fitting our rheological curves.

With the yield-stress ($\sigma_y$) obtained from Fig. 2 (of the main text), the other free parameters are the time-scale $\tau$ and the pre-factor $\eta$ in the Bagnold-branch, $\sigma^{(\text{rep})} = \eta\dot{\gamma}^2$. The value of $\eta$ only depends on $\phi$ [7]. The time-scale $\tau$ follows from Fig.4b as $\tau = c/u$. The constant $c \approx 3.5$ is independent of $u$ and $\phi$.

---